\documentclass{JHEP3} 

\JHEPspecialurl{http://jhep.sissa.it/JOURNAL/JHEP3.tar.gz}
\usepackage{amsmath}
\usepackage{multirow}
\usepackage{booktabs}
\usepackage{multirow}
\newcommand\ee{\end{equation}}
\newcommand\be{\begin{equation}}
\newcommand\eea{\end{eqnarray}}
\newcommand\bea{\begin{eqnarray}}

\def\p{\pi}

\def\th{\theta}

\def\s{\sigma}

\def\s{\sigma}
\def\nn{\nonumber}
\def\p{\partial}

\newcommand{\bi}{\begin{itemize}}
\newcommand{\ei}{\end{itemize}}
\renewcommand{\th}{\theta}
\newcommand{\bth}{\bar{\theta}}

\newcommand\fverb{\setbox\fverbbox=\hbox\bgroup\verb}
\newcommand\fverbdo{\egroup\medskip\noindent%
            \fbox{\unhbox\fverbbox}\ }
\newcommand\fverbit{\egroup\item[\fbox{\unhbox\fverbbox}]}
\newbox\fverbbox

\title{On Ghost-free Supersymmetric Galileons}

\author{Fotis Farakos$^a$, Cristiano Germani$^b$ and Alex Kehagias$^{a,c}$
\\
$^a$\!\!
   Physics Division, National Technical University of Athens, \\15780 Zografou Campus, Athens, Greece\\
${}^b$\!\! Arnold Sommerfeld Center, Ludwig-Maximilians-University,\\
Theresienstr. 37, 80333 Muenchen, Germany\\
${}^c$ \!\!Universit\'e de Gen\`eve, D\'epartement de Physique\\
Th\'eorique and CAP, 24 quai Ernest-Ansermet, \\
CH-1211, Gen\`eve 4, Switzerland\\
E-mail: \email{fotisf@mail.ntua.gr, cristiano.germani@lmu.de,  kehagias@central.ntua.gr}}

\abstract{
We present consistent supersymmetric theories invariant under the generalization of the Galilean shift symmetry to ${\cal{N}}=1$ superspace. 
These theories are constructed via the decoupling limit of certain non-minimally derivative coupled supergravities, thus they correspond to the supersymmetrization 
of the so-called covariant Galileon. Specifically, these theories are constructed in the linearized ${\cal{N}}=1$ new-minimal supergravity set-up where the chiral supermultiplet is minimally coupled to gravity via the standard R-current contact term, and, at the same time, non-minimally derivatively coupled to the Einstein superfield. }

\begin{document}


\section{Introduction}

It has been recently discovered \cite{non,non2} that there is a set of, non-renormalizable, scalar field self-interactions having the interesting property that their suppression scale do not run at any energies. In addition, these same theories enjoy, in flat space, the following Galilean shift
\be\label{shift}
\pi\rightarrow \pi+c+b_\mu x^\mu\ ,
\ee
where $c,b_\mu$ are respectively a constant and a constant four-vector, $x^\mu$ are Cartesian coordinates in a Minkowski spacetime and $\pi$ is the so-called {\it Galilean} 
field \cite{fullset,Padilla:2010de}. The additional requirement to have only up to second order differential equations, in order to avoid possible Ostrogradski instabilities, restrict the Galilean Lagrangians to contain only a product of up to five scalars \cite{fullset}.

The transformation \eqref{shift} may be extended to superspace. 
In particular, for a  chiral superfield ($\Phi$), 
we propose that the only consistent supersymmetric extension of \eqref{shift} is
\be\label{superspaceG0}
\Phi  \rightarrow \Phi + c + b_\mu  y^\mu
\ee
where
\be
\label{yintr}
y^\mu =x^\mu  + i \theta \s^\mu  \bar{\theta}\ .
\ee
Note that the super-Galilean shift \eqref{superspaceG0}, when projected to the real space, only shifts the lowest component of $\Phi$ as the complex extension of \eqref{shift} while it maintains its superspace chirality.

It has been shown in \cite{Koehn:2013hk}, by brute-force calculation, that the supersymmetrization of a cubic Galileon out of a chiral field \cite{Khoury:2011da}, is not possible without the appearance of ghosts. The same Authors however, could not exclude the quartic supersymmetric Galilean theories, although their constructions only led to ghost-propagating field theories. Those theories, although invariant under \eqref{shift}, were not invariant under the superspace Galilean shift \eqref{superspaceG0} introduced here. We believe, that this was the main issue that led the Authors of \cite{Koehn:2013hk} to conclude that no supersymmetric Galileons can be found without propagating ghosts states.

In this paper we indeed show that a ghost-free quartic Galilean theory {\it does} exist and is invariant under \eqref{superspaceG0}. In other words, we will construct the supersymmetric version of the so-called ``quadratic" and ``quartic" Galileons
\bea\label{flat}
{\cal L}_2&=&-\frac{1}{2}\partial_\mu\bar\pi\partial^\mu\pi\ ,\cr
{\cal L}_4&=&- \frac{4}{\Lambda^{6}}
\pi   (\p_{[\kappa } \p^\kappa  \bar{\pi})(\p_\lambda  \p^\lambda  \pi)(\p_{\zeta ]} \p^\zeta  \bar{\pi})\ ,
\eea
where $\bar\pi$ is the complex conjugated of $\pi$ and $\Lambda$ is a suppression scale.

The easiest way to find our quartic Galilean theory passes through a decoupling limit of certain supergravities. To appreciate this, let us go back to the non-supersymmetric case.

In Minkowski space, the shift $b_\mu x^\mu=b_a \int \xi^a_\mu dx^\mu$, where $\xi^a_\mu=\delta^a_\mu$ is a set of Killing vectors (the four related to translations and labelled by $a$) such that $\nabla_\mu\xi_\nu^a=0$ (i.e. integrable), and $b_a=\delta_a^\mu b_\mu$. One may then ask the question of whether generalized ``Galilean" theories, i.e. with the property that they are invariant under the shift
\be
\pi\rightarrow\pi+c+b_a\int \xi^a_\mu dx^\mu\ ,
\ee
exist in non-trivial spacetimes with integrable Killing vectors. 

This question has been answered in \cite{sloth}. In particular, up to quadratic order in $\pi$ one has 
\bea
\label{curved}
{\cal A }_{2}&=&-\frac{1}{2}g^{\mu\nu}\partial_\mu\pi\partial_\nu\bar\pi+\frac{1}{2 M^2}G^{\mu\nu}\partial_\mu\pi\partial_\nu\bar\pi\ .
\eea
where $M$ is a mass scale and $G^{\mu\nu}$ is the Einstein tensor. 
The sign of the terms in ${\cal A}_2$ are chosen in such a way that, whenever energy conditions are satisfied, the effective propagator of $\pi$ is never ghost-like \cite{new}. 

Again the theories of \cite{sloth} enjoy a non-renormalization theorem (up to the Planck scale) of their coupling/suppression constants \cite{sloth}. Finally the theory ${\cal A}_2$ has been dubbed {\it Slotheonic} theory in \cite{sloth} (and so $\pi$ the ``{\it Slotheon}") for its property of a ``slow" scalar evolution with respect to the minimal case $M\rightarrow\infty$. This property, turned out to be the key issue to produce successful inflationary scenarios even in the case of steep scalar field potentials \cite{new, uv}.

Thanks to the equivalence principle, locally, any spacetime is approximately flat. Another way to see this is to notice that, in Riemanian coordinates, for any theory where graviton self-interaction is suppressed by the Planck scale, 
\be
\nabla_\alpha\xi_\beta={\cal O}(\frac{1}{M_p})\ ,
\ee 
where $\partial_\mu\xi_\nu=0$ and $M_p$ is the Planck scale. 
Therefore, there must exist ``decoupling limits" involving $M_p\rightarrow \infty$, such that (\ref{curved}), endowed with the Einstein-Hilbert Lagrangian 
\be
{\cal L} _{\rm grav}=\frac{1}{2}M_p^2 R\ ,
\ee
reproduces (\ref{flat}). 

These limits have been found in \cite{decoupling} showing an intimate relation between the theories (\ref{flat}) and (\ref{curved}), i.e. between Galileons and Slotheons. In particular, in \cite{decoupling}, it has been shown that the Lagrangian
\be
{\cal L}=\frac{1}{2}\left[M_p^2 R-\frac{1}{M^2}G^{\mu\nu}\partial_\mu\pi\partial_\nu\pi\right]\ ,
\ee
in the limit $M_p\rightarrow\infty$ but $\Lambda=M^2 M_p\rightarrow {\rm finite}$, reproduces the quartic Galileon ${\cal L}_4$. In the non-decoupling limit instead, with the help of the gravity equations, the equation of motion for $\pi$ are nothing else than the covariant Galileon of \cite{covariant}.

\section{Non-minimally kinetically coupled Supergravity}

Following \cite{FGKS}, we will work in the ${\cal{N}}=1$ new-minimal supergravity framework \cite{newminimal}. 
Apart from \cite{FGKS}, higher derivative extensions of new-minimal supergravity have been also 
studied in \cite{sugra-hd}, whereas consistent higher derivative theories have been discussed in \cite{ovrut,kf1}.

As we will only be interested in the decoupling limit of gravity, 
we will only consider Lagrangians at linearized level in the graviton \cite{Cecotti:1987qe}. The non-minimal derivative coupling
of a chiral superfield  $\Phi$ to the linearized new-minimal supergravity, is found by considering the supersymmetric lagrangian \cite{FGKS} 
\be
\label{L0}
{\cal L }_{0}= \int d^4\theta  \frac{2 i}{M^{2}}   \Phi E^\mu  \p_\mu   \bar{ \Phi}    
\ee
where $E^\mu $ is the Einstein superfield. 
We recall that the Einstein superfield is defined 
in terms of the real superfield $\phi_\mu $ as 
\be
\label{E}
E^\mu =-\frac{1}{2} \epsilon^{\mu\nu\rho\sigma} \bar{D}_{\dot{\alpha}} \bar{\s}_{\nu}^{\dot{\alpha}  \alpha}  
D_{\alpha} \p_\rho \phi_\s
\ee
where the covariant derivatives with respect to the Grassman co-ordinates of the superspace are defined as usual
\be
D_{\alpha}=\frac{\p}{\p \theta^{\alpha}} + i \s_{\alpha \dot{\alpha}}^\mu  \bar{\theta}^{\dot{\alpha}} \p_\mu  \ , 
\ \ 
\label{D's}
\bar{D}_{\dot{\alpha}}=-\frac{\p}{\p \bar{\theta}^{\dot{\alpha}}} - i \theta^{\alpha} \s_{\alpha \dot{\alpha}}^\mu   \p_\mu. 
\ee
The real superfield $\phi_\mu$ is invariant under the following gauge symmetry (needed in order 
to contain the $12+12$ degrees of freedom of new-minimal supergravity)
\be
\label{phi-gauge}
\delta \phi_\mu = \p_\mu V +S_\mu + \bar{S}_\mu\ ,
\ee
with $V$ a real superfield and $S_\mu$ a chiral superfield. 

Obviously, the superfield $E_\mu$ is also invariant under this gauge transformation. In fact, $E_\mu$ is nothing else than the ``field strength`` of $\phi_\mu$.

In the appropriate Wess-Zumino (WZ) gauge, $\phi_\mu$ contains the graviton $h_{\mu\nu}$, the gravitino $\psi_\mu$,
a two-form auxiliary $B_{\mu\nu}$ and a vector auxiliary $A_\mu$. The latter, gauges the continuous R-symmetry in supergravity.

The $\th$-expansion of $\phi_\mu$ is explicitly written as
\be
\phi_\mu|_{_{WZ}}=-\th\s^\nu\bth(h_{\nu\mu}+B_{\nu\mu})
+i\theta^2\bth \bar{\psi}_\mu 
-i\bth^2\theta \psi_\mu
+\frac{1}{2}\th^2\bth^2A_\mu \label{f}
\ee
and it is useful to define the components of $\phi_\mu$ in terms of projections as \footnote{As standard we use the notation ``$|$" to mean $\theta=\bar\theta=0$. For our superspace
conventions see \cite{SUGRA}.} 
\begin{align}
-\frac{1}{2} [D_{\alpha}, \bar{D}_{\dot{\alpha}}] \phi_\mu  | &= h_{\alpha \dot{\alpha} \mu} + 
B_{\alpha \dot{\alpha} \mu}
\\
\nn
- \frac{i}{4} \bar{D}^2 D_{\alpha}   \phi_\mu  | &= \psi_{\mu \alpha}
\\
\nn
\frac{i}{4} D^2 \bar{D}_{\dot{\alpha}}   \phi_\mu  | &= \bar{\psi}_{\mu \dot{\alpha}}
\\
\label{wz-comp}
-\frac{1}{8} D^{\alpha}   \bar{D}^2  D_{\alpha}  \phi_\mu  |& =  A_\mu. 
\end{align}
Using (\ref{f}) in (\ref{E}), we find that $E_\mu$ can be expanded as
\begin{align}
E_\mu&=-2H_\mu-2 i \th r_\mu+2i\bth \bar{r}_\mu -\th \s^\nu \bth (G_{\nu\mu}+\p^\lambda H_{\lambda\nu\mu}
-\,^*F_{\nu\mu})\nonumber \\
&+\th^2\bth \bar{\s}^\nu\p_\nu r_\mu
-\bth^2\th \s^\nu\p_\nu \bar{r}_\mu
-\frac{1}{2}\th^2\bth^2\p^2H_\mu
\end{align}
where 
\begin{align}
\nn
H^\mu&=\frac{1}{3!}\epsilon^{\mu\nu\rho\sigma}H_{\nu\rho\sigma}\, , \\
H_{\nu\rho\s} &= (\p_\nu B_{\rho\s} +  \p_\rho B_{\s\nu} +\p_\s B_{\nu\rho} ) \nn
\\
\nn
\ ^* F^{\mu\nu} &= \frac{1}{2} \epsilon^{\mu\nu\rho\s} ( \p_\rho A_\s - \p_\s A_\rho )
\label{def's}
\end{align}
and 
\begin{align}
R_{\mu\rho} &= - \p_\mu  \p_\rho h^{\nu}_{\nu}  +  \p^\nu \p_\rho h _{\nu\mu} +   \p^\nu \p_\mu  h _{\nu\rho} -  
\p^2 h _{\mu\rho}\ ,\nn\\
G_{\mu\nu}&=R_{\mu\nu}-\frac{1}{2}\eta_{\mu\nu}R
\end{align}
are the linearized Ricci and Einstein tensors respectively.

The components of $E^\mu $ can be found using the definitions (\ref{wz-comp}) and the supersymmetry algebra
\begin{align}
\nn
-\frac{1}{2} E^\mu  | &= H^\mu  
= \frac{1}{3!} \epsilon^{\mu\nu\rho\s} H_{\nu\rho\s}\ , 
\\
\nn
\frac{i}{2}  D_{\alpha}   E^\mu  | &= r^{\mu}_{ \alpha} 
= - \frac{1}{2} \epsilon^{\mu\nu\rho\s} \s_{\nu \alpha \dot{\alpha}} \p_\rho  \bar{\psi}_{\s}^{\dot{\alpha}}\ , 
\\
\label{E-comp}
-\frac{1}{2} [D_{\alpha}, \bar{D}_{\dot{\alpha}}] E_\mu  | &= \s^{\nu}_{\alpha \dot{\alpha}} 
( E_{\nu \mu} 
+ \p^\lambda H_{\lambda \nu\mu} 
- \ ^*F_{\nu\mu} ).
\end{align}
Note that $r^{\mu}_{ \alpha}$ is the field strength of the Rarita-Schwinger field, 
moreover,  $E^\mu $ is a real linear superfield as it satisfies the conditions
\be
\bar{E}^\mu =E^\mu \,, ~~~D^2E^\mu =0
\ee
as well as the superspace Bianchi identity
\be
\partial_\mu E^\mu =0.
\ee
The components of the chiral superfield are defined as usual
\bea
\nn
\Phi| &=&\pi
\\
\nn
\frac{1}{\sqrt{2}} D_{\alpha} \Phi|&=&\chi_\alpha 
\\
\label{chiral-def}
- \frac{1}{4}  D^2 \Phi| &=&F.
\eea

Taking into account the standard coupling of the Einstein superfield $E^\mu$ with the graviton multiplet 
$\phi_\mu$, we can write the leading terms of a chiral superfield $\Phi$ coupled  to the new-minimal 
linearized supergravity as 
\be
\label{L1}
{\cal L }_{1}= \int d^4\theta \left( M_P^2  E^\mu  \phi_\mu  +  \bar{\Phi} \Phi +\phi^\mu R_\mu 
- \frac{2 i}{M^{2}}   \Phi E^\mu  \p_\mu   \bar{ \Phi}    \right) \, +\,{\cal{O}}\left(\frac{1}{M_P}\right)\ .
\ee
In addition,  $R_\mu$ is the supersymmetric R-current (see for example \cite{supercurrents})
which is defined as
\be 
R_\mu= - \bar{\s}_{\mu}^{ \dot{\alpha}  \alpha} D_{\alpha} \Phi \bar{D}_{\dot{\alpha}} \bar{\Phi}
\ee
and satisfies (on-shell)
\be
\bar{D}^{\dot{\alpha}}R_{\alpha \dot{\alpha}}=\chi_{\alpha}
\ee
with
\be 
\chi_{\alpha}= \bar{D}^2 D_{\alpha} (\bar{\Phi} \Phi).
\ee
Note that, in the spirit of the already mentioned decoupling limit, we have silently assumed that $M$ is not proportional to 
$M_P$ (in which case we could have omitted the term (\ref{L0}) from (\ref{L1})) 
but rather, as we will see later, proportional to $1/M_P^{1/2}$.

Concerning dimensions, we have assigned mass dimension zero to the graviton 
but the graviton multiplet has $[\phi_\mu ]=-1$.
For the chiral superfield $[\Phi]=1$ and for the superspace derivatives 
$([D_{\alpha}][\bar{D}_{ \dot{\alpha}}])\sim[\p_\mu ]=1$.

\subsection{Decoupling limit}
We now proceed to the decoupling of gravity as in the previous discussions and \cite{decoupling}. 

The (gravity) equations of motion for $\phi_\mu$ are
\be
\label{grav-EOM}
 E^\s  + \frac{1}{2M_P^2} R^\s
+ \frac{i}{2M_P^2 M^{2}} \bar{D}_{\dot{\alpha}} \p_\mu  \bar{\Phi} \bar{\s}_{\nu}^{\dot{\alpha}  \alpha} 
D_{\alpha} \p_\rho \Phi  
\epsilon^{\mu\nu\rho\s} =0
\ee
and for $\Phi$ we have
\be
\label{matter-EOM}
\bar{D}^2( \bar{\Phi} 
-  \bar{\s}_{\mu}^{\dot{\alpha} \alpha} D_{\alpha} (\phi^{\mu} \bar{D}_{\dot{\alpha}} \bar{\Phi} ) 
-2i  \frac{1}{ M^{2}} E^\mu  \p_\mu  \bar{\Phi}) =0.
\ee
Solving for $E^\mu $ in (\ref{grav-EOM}) and plugging into (\ref{matter-EOM}) we find
\begin{align}
\nn
 \bar{D}^2( \bar{\Phi} 
&-\bar{\s}_{\mu}^{\dot{\alpha} \alpha} D_{\alpha} (\phi^{\mu} \bar{D}_{\dot{\alpha}} \bar{\Phi} ) 
- \frac{i}{ M^2 M_P^2} R^\mu \p_\mu  \bar{\Phi}
\\
\label{combined}
&-\frac{1}{M_P^2 M^{4}}   
(\bar{D}_{\dot{\alpha}} \p_\mu  \bar{\Phi} \bar{\s}_{\nu}^{\dot{\alpha}  \alpha} D_{\alpha} \p_\rho \Phi )
\epsilon^{\mu\nu\rho\s}  \p_\s \bar{\Phi})
 =0.
\end{align}
Now, in the limit
\be
M_P\to \infty, ~~\mbox{such that}~~~M^2M_P=\Lambda^3 ~~~ \mbox{is finite}
\ee
gravity decouples 
\be
E_\mu =0 ~~ \rightarrow ~~ \phi_\mu = \mbox{pure gauge} 
\ee
and 
\be
\label{combined1}
 \bar{D}^2( \bar{\Phi} 
-  \frac{1}{\Lambda^6}   
(\bar{D}_{\dot{\alpha}} \p_\mu  \bar{\Phi} \bar{\s}_{\nu}^{\dot{\alpha}  \alpha} D_{\alpha} \p_\rho \Phi )
\epsilon^{\mu \nu \rho \s}  \p_\s \bar{\Phi})
 =0
\ee
where in (\ref{combined1}), using the fact that $\phi_{\mu}$ is a pure gauge, we have set it to zero.

The component form of (\ref{combined1}) (ignoring all fermions 
and auxiliary fields) are 
\be
\label{combined-components}
\p^2 \bar{\pi} 
-\frac{4}{\Lambda^6} (\p_{[\kappa} \p^\kappa \bar{\pi})(\p_\lambda \p^\lambda \pi)(\p_{\zeta]} \p^\zeta \bar{\pi}) = 0\ ,
\ee
which is just the complex Galilean equation of motion coming from the variation of the action (\ref{flat}), as anticipated.

\subsection{Supersymmetric Galileon}
Now that we learned the structure of the quartic supersymmetric Galileon as a decoupling limit of the new non-minimally coupled ${\cal N}=1$ supergravity of \cite{FGKS}, we can infer the superspace Lagrangian 
that gives rise to the superspace equations of motion (\ref{combined1}). 

After a straightforward calculation one then finds that the Lagrangian describing the super-galileon is given by
\be
\label{L2}
{\cal L }= \int d^4\theta (\bar{\Phi} \Phi
-\frac{1}{ \Lambda^{6}}\Phi
(\bar{D}_{\dot{\alpha}} \p_\mu  \bar{\Phi} \bar{\s}_{\nu}^{\dot{\alpha}  \alpha} D_{\alpha} \p_\rho \Phi )
\epsilon^{\mu \nu \rho \s}  \p_\s \bar{\Phi}).
\ee

The Lagrangian (\ref{L2}), on top of the standard supersymmetries, enjoys the galilean symmetry extended to superspace, i.e.
\begin{align}
\Phi  &\rightarrow \Phi + c + b_\mu  y^\mu\label{supergal}
\end{align}
where $c$ is a complex constant, $b_\mu $ is a complex constant vector and 
\be
\label{y}
y^\mu =x^\mu  + i \theta \s^\mu  \bar{\theta}.
\ee
The latter satisfies  the relations
\be
\label{y-rel}
\bar{D}_{\dot{\alpha}} y_\nu = 0 , 
\ \ D_{\alpha} y_\nu= 2i \s_{\nu \alpha \dot{\alpha}} \bar{\theta}^{\dot{\alpha}} , 
\ \ D^2 y_\nu = 0, \ \ \p_\mu  y_\nu= \eta_{\mu\nu}.
\ee
The super-galilean symmetry (\ref{supergal}) is defined in a way such that:
\begin{itemize}

\item   it preserves the chirality  of the superfield $\Phi$ ($\bar{D}_{\dot{\alpha}}\Phi=0$)

\item   it induces  the following galileon transformations for the scalar ($\pi$), its fermionic super-partner ($\chi_\alpha$) and the auxiliary field ($F$)
\bea\label{long}
\pi&\rightarrow&\pi+c+b_\mu x^\mu\ ,\nn\cr
\chi_\alpha&\rightarrow& \chi_\alpha\nn\ ,\cr
F&\rightarrow& F\ .
\eea
\end{itemize}
Note that one could certainty start from requiring the symmetries (\ref{supergal}) to
obtain (via a series of trial and errors) the action (\ref{L2}), without ever
invoking the decoupling limit. Nevertheless, we find the decoupling way somehow
more fundamental. Indeed,
in this way, one has automatically at hand the supergravity extension of the
quartic Galileon. 

The component form of (\ref{L2}) is 
\bea
\nn
{\cal L }&=& \pi  \p^2 \bar{\pi }  + i \p_\nu \bar{\chi}  \bar{\s}^\nu \chi + F \bar{F}
\\
\nn
&-& \frac{1}{\Lambda^{6}}  (
4 \pi   (\p_{[\kappa } \p^\kappa  \bar{\pi})(\p_\lambda  \p^\lambda  \pi)(\p_{\zeta ]} \p^\zeta  \bar{\pi}) 
 -8F \p_\mu \bar{F} \p_\s  \bar{\chi} \bar{\s}_\nu  \p_\rho  \chi \epsilon^{\mu \nu \rho \s } 
\\
\nn
&& \ \ \ \ \ 
-4i \p_\mu  \chi \s^\tau  \p_\tau  \bar{\chi} \p_\s  \bar{\chi} \bar{\s}_\nu \p_\rho  \chi \epsilon^{\mu \nu \rho \s }
-2i \p_\mu  \bar{\chi} \bar{\s}_\nu \s^\kappa  \p_\s  \bar{\chi }  \epsilon^{\mu \nu \rho \s }  \p_\kappa  \chi \p _\rho  \chi
\\
\nn
&& \ \ \ \ \   
+4i \p_\s  \chi \s^\kappa  \bar{\s}_\nu \p_\rho  \chi \epsilon^{\mu \nu \rho \s } \p_\mu  \bar{F} \p_\kappa  \bar{\pi}
+4i\p_\rho  F \p_\mu  \bar{\chi} \bar{\s}_\nu \s^\kappa  \p_\s  \bar{\chi } \epsilon^{\mu \nu \rho \s } \p_\kappa    \pi 
\\
\nn
&& \ \ \ \ \    
+4 \p_\mu  \pi \p^2 \bar{\pi} \epsilon^{\mu \nu \rho \s } \p_\s  \bar{\chi} \bar{\s}_\nu \p_\rho  \chi
+ 2 \p_\mu  \bar{\chi} \bar{\s}_\nu \s^\lambda  \bar{\s}^\kappa  \p_\lambda  \chi  \epsilon^{\mu \nu \rho \s }   \p_\rho  \pi   \p_\kappa  \p_\s  \bar{\pi}
\\
\nn
&& \ \ \ \ \ +  4 \p_\tau  \bar{\chi} \bar{\s}^\tau \s^\kappa  \bar{\s}^\nu \p_\rho  \chi \epsilon^{\mu \nu \rho \s } \p_\mu \pi \p_\kappa  \p_\s  \bar{\pi} 
+2 \chi \s^\kappa  \bar{\s}_\nu \s^\lambda  \p_\s  \bar{\chi} \epsilon^{\mu \nu \rho \s } \p_\kappa  \p_\mu  \bar{\pi} \p_\lambda  \p_\rho  \pi
\\
\label{LT2-components}
&& \ \ \ \ \   
+2  \p_\mu  \bar{\chi} \bar{\s}_\nu \s^\kappa  \bar{\s}^\lambda  \chi \epsilon^{\mu \nu \rho \s } \p_\lambda  \p_\s  \bar{\pi} \p_\kappa  \p_\rho  \pi
-2 \p_\lambda  \chi \s^\kappa    \bar{\s}_\nu  \s^\lambda   \p_\s  \bar{\chi}  \epsilon^{\mu \nu \rho \s } \p_\mu  \p_\kappa  \bar{\pi}  \p_\rho  \pi
 ).
\eea
In order to find the final Lagrangian, one should integrate out the auxiliary field $F$ in \eqref{LT2-components}. The way to do that closely resemble the case studied in \cite{seiberg}.

Variation of \eqref{long} with respect to $\bar F$ gives schematically an equation of type
\be\label{Feq}
F+\frac{\alpha^\mu}{\Lambda^6}\partial_\mu F +\frac{\beta}{\Lambda^6}=0\ ,
\ee
where $\alpha^\mu$ and $\beta$ are functions of the scalar field $\pi$ but most importantly of the fermionic field $\chi$. Finally, the scale $\Lambda$ has been explicitly extracted. To solve \eqref{Feq} one can use an iterative procedure. The first step is to invert this equation as
\be
F=-\frac{\alpha^\mu}{\Lambda^6}\partial_\mu F -\frac{\beta}{\Lambda^6}\ .
\ee
The second step would be to substitute again the inversion, i.e.,
\be
F=-\frac{\alpha^\mu}{\Lambda^6}\partial_\mu\left(-\frac{\alpha^\nu}{\Lambda^6}\partial_\nu F -\frac{\beta}{\Lambda^6}\right)-\frac{\beta}{\Lambda^6}\ ,
\ee
and so on. Thanks to the Grassmanian properties of the fermions $\chi$ this recursion eventually ends as soon as more than two equal fermions are multiplied (this is typical in supersymmetric theories, see for example \cite{seiberg}). 
The final Lagrangian is very involved and not very enlightening, for this reason we leave the interested reader to do the full inversion. Nevertheless, as the cut-off of the theory is $\Lambda$, which also corresponds to the suppression scale of the pure Galilean term, it is interesting to consider the supersymmetric action \eqref{LT2-components} up to ${\cal O}(\Lambda^{-12})$.
Equation \eqref{Feq} is solved for
\be
F={\cal O}(\Lambda^{-6})\ .
\ee
Therefore the supersymmetric Galilean action to leading order in the cut-off scale $\Lambda$ reads
\be\label{completa}
{\cal L }_{\rm gal}^{\tiny (\Lambda)}={\cal L }_{\rm WZ}+
\frac{1}{\Lambda^6}\left[{\cal L }_{\pi\pi}+{\cal L }^{\tiny (0)}_{\pi\chi}\right]
\ee
where
\be
{\cal L }_{\rm WZ}=\pi  \p^2 \bar{\pi }  + i \p_\nu \bar{\chi}  \bar{\s}^\nu \chi
\ee
is the Wess-Zumino action,
\be
{\cal L }_{\pi\pi}=-4 \pi   (\p_{[\kappa } \p^\kappa  \bar{\pi})(\p_\lambda  \p^\lambda  \pi)(\p_{\zeta ]} \p^\zeta  \bar{\pi}) 
\ee
is the scalar Galilean self-interaction, and finally the mix fermion-scalar interaction Lagrangian is
\begin{align}\nn
{\cal L }^{\tiny (0)}_{\pi\chi}=
&-4i \p_\mu  \chi \s^\tau  \p_\tau  \bar{\chi} \p_\s  \bar{\chi} \bar{\s}_\nu \p_\rho  \chi \epsilon^{\mu \nu \rho \s }
-2i \p_\mu  \bar{\chi} \bar{\s}_\nu \s^\kappa  \p_\s  \bar{\chi }  \epsilon^{\mu \nu \rho \s }  \p_\kappa  \chi \p _\rho  \chi
\\
\label{pc}
&  
+4 \p_\mu  \pi \p^2 \bar{\pi} \epsilon^{\mu \nu \rho \s } \p_\s  \bar{\chi} \bar{\s}_\nu \p_\rho  \chi
+ 2 \p_\mu  \bar{\chi} \bar{\s}_\nu \s^\lambda  \bar{\s}^\kappa  \p_\lambda  \chi  \epsilon^{\mu \nu \rho \s }   \p_\rho  \pi   \p_\kappa  \p_\s  \bar{\pi}
\\
\nn
& +  4 \p_\tau  \bar{\chi} \bar{\s}^\tau \s^\kappa  \bar{\s}^\nu \p_\rho  \chi \epsilon^{\mu \nu \rho \s } \p_\mu \pi \p_\kappa  \p_\s  \bar{\pi} 
+2 \chi \s^\kappa  \bar{\s}_\nu \s^\lambda  \p_\s  \bar{\chi} \epsilon^{\mu \nu \rho \s } \p_\kappa  \p_\mu  \bar{\pi} \p_\lambda  \p_\rho  \pi
\\
& 
+2\p_\mu  \bar{\chi} \bar{\s}_\nu \s^\kappa  \bar{\s}^\lambda  \chi \epsilon^{\mu \nu \rho \s } \p_\lambda  \p_\s  \bar{\pi} \p_\kappa  \p_\rho  \pi
-2 \p_\lambda  \chi \s^\kappa    \bar{\s}_\nu  \s^\lambda   
\p_\s  \bar{\chi}  \epsilon^{\mu \nu \rho \s } \p_\mu  \p_\kappa  \bar{\pi}  \p_\rho  \pi\ .\nn
\end{align}
Note that, from \eqref{Feq}, the full ${\cal L }_{\rm gal}$, i.e. at all orders in $\Lambda$, would only involve extra $\pi$, $\chi$ interaction terms suppressed by higher powers of the cut-off scale $\Lambda$. In other words, the full Galilean action would only modify eq. \eqref{pc} by additional ${\cal O}(\Lambda^{-6})$ terms. Explicitly
\be\nn
{\cal L }_{\rm gal}={\cal L }_{\rm WZ}+
\frac{1}{\Lambda^6}\left[{\cal L }_{\pi\pi}+{\cal L }_{\pi\chi}\right]
\ee
where
\be\nn
{\cal L }_{\pi\chi}={\cal L }_{\pi\chi}^{\tiny (0)}+{\cal O}\left(\frac{1}{\Lambda^6}\right)\ .
\ee

The supergravity action  \eqref{L1}, does not contain higher derivatives of either bosons or fermions. It just 
describes a complex scalar, a fermion, a graviton and a gravitino. 
It is clear then that no new fields may emerge in the decoupling limit. 
In fact is quite the contrary. By definition, in the decoupling limit some degrees of freedom (the whole gravity multiplet) are ``lost". Of course, they reappear at the scale $\Lambda$. Thus, at the perturbative level, for energies
below $\Lambda$, our theory is well described by a complex scalar field and a fermion (the scalar multiplet). This is the power of using the decoupling limit.

We can see this even from another perspective:
It is clear that the bosonic sector
contains only a complex scalar and no extra ghost states, in accordance with the Ostrogradsky theorem. 
Exact supersymmetry does not allow for extra states behind a single fermionic one, the superpartner of the scalar galileon.
An extra (ghost) fermionic state cannot be present in the spectrum as it cannot be paired with a corresponding 
(ghost) bosonic state to form a scalar ghost supermultiplet, simply because such bosonic state does not exist.
Summarizing, as there is no bosonic ghost state to pair with, there is no ghost supermultiplet either and therefore no 
any fundamental ghost states at all.   

\section{Conclusions}

Galilean invariant theories have attracted a huge attention lately. One of the most striking properties is that their suppression scales do not (quantum mechanically) run with energy. 

Using the superspace formalism, one would already guess that, if the projected theory onto the real space should be Galilean invariant, in superspace, this symmetry must be incorporated into a larger symmetry. 

Indeed, we showed that a Galilean theory must be embedded into a super-space Lagrangian (i.e. before projection to real space) enjoying the super-space Galilean shift symmetry
\be\label{superspaceG}
\Phi  \rightarrow \Phi + c + b_\mu  y^\mu
\ee
where
\be
\label{yconcl}
y^\mu =x^\mu  + i \theta \s^\mu  \bar{\theta}\ ,
\ee
and where $\Phi$ is the Galilean chiral superfield. Note that the super-Galilean shift \eqref{superspaceG}, in components, only shifts the scalar $\pi$. 

The way we found our supersymmetric Galilean Lagrangian was however somehow indirect.

Inspired by the result of \cite{decoupling} showing that the complex Galilean Lagrangians may be found as a decoupling limit ($M_p\rightarrow\infty$ but $\Lambda=M^2M_p$ finite) of
\be\label{slothsugra}
{\cal L}_{\rm slotheon} =\frac{1}{2}\left[M_p^2 R-\frac{G^{\mu\nu}}{2M^2}\partial_\mu\pi\partial_\nu\bar\pi\right]\ ,
\ee
we used the supergravity extension of the theory \eqref{slothsugra} found in \cite{FGKS} to obtain our supersymmetric Galilean Lagrangian \eqref{LT2-components}. Thus, the theory \cite{FGKS} is the supergravity extension of Galilean theories, i.e. the covariant super-Galilean theory.

We would like to conclude by noticing that the theory \cite{FGKS} could only be found in the new-minimal supergravity formalism which requires a conservation of R-charge. In particular it turned out that the chiral superfield could only have vanishing R-charge. In the decoupling limit this is perfectly consistent with the Galilean shift \eqref{superspaceG}. In fact, the super Galilean shift has vanishing R-charge and therefore it can only be applied to superfields with vanishing R-charge as well. 
Thus, in order to have a consistent R-invariant theory, the super-Galileon must have vanishing R-charge, as required by the supergravity extension.

This observation may also be related to the statement of \cite{Koehn:2013hk} that 
cubic super-Galilean theories cannot be constructed out of chiral superfields. 
In \cite{decoupling}, it has been shown that the cubic Galilean theory can be obtained as a decoupling limit of a 
theory containing both the ``Slotheonic door" $G^{\mu\nu}\partial_\mu\pi\partial_\nu\pi$ and the conformal 
coupling $\pi R$. However, it turns out that the two terms cannot coexist in the new-minimal 
supergravity formalism, as, the first would require a vanishing R-charge contrary to the second. 
Thus, the cubic super-Galileon cannot be obtained as a decoupling limit of the new-minimal
supergravity theory coupled to chiral superfields, as has been done here for the quartic galileon. 
The quintic Galileon is instead more mysterious. In \cite{decoupling} no consistent decoupling limit has been found 
such to lead to the quintic Galileon. Therefore, its supersymmetrization cannot procceed along the lines followed here
for the quartic galileon. 

\section*{Acknowledgements}

The authors thank Jean-Luc Lehners and Ulf Lindstr\"om for discussion and
correspondence.
This research was implemented under the ``ARISTEIA" Action of the
``Operational Programme Education and Lifelong Learning''
and is co-funded by the European
Social Fund (ESF) and National Resources.  It is partially
supported by European Unions Seventh Framework Programme (FP7/2007-2013)
under REA
grant agreement n. 329083.
CG is supported by Humboldt foundation.

\appendix

\end{document}